\documentclass[prd,showpacs,twocolumn,nofootinbib,preprintnumbers,amsmath,amssymb]{revtex4-1}
\pdfoutput=1

\usepackage{amsmath,amssymb,url}
\usepackage{amssymb,amsfonts,multirow}
\usepackage{rotating}
\usepackage{color}
\usepackage{hhline}
\usepackage{slashed}

\usepackage{xcolor}
\usepackage{multirow}

\usepackage{array,multirow}

\usepackage{booktabs}

\definecolor{Gray}{gray}{0.85}
\definecolor{LightGreen}{rgb}{0.88, 1, 0.88}
\definecolor{Lime}{rgb}{0,255,0}
\definecolor{LightCyan}{rgb}{0.88,1,1}
\definecolor{LightRed}{rgb}{1, 0.85, 0.85}
\definecolor{Red}{rgb}{1, 0, 0}
\definecolor{LightYellow}{rgb}{1, 1, 0.85}
\definecolor{Yellow}{rgb}{1,1,0.05}
\definecolor{LightBlue}{rgb}{0.87, 0.94, 1}
\definecolor{white}{gray}{1}
\definecolor{black}{gray}{0}
\newcolumntype{G}{>{\columncolor{LightGray}}c}

\usepackage{framed}
\usepackage{bm}
\usepackage{amsmath}
\usepackage{graphicx}
\usepackage{amssymb}
\usepackage{epstopdf}
\usepackage{hyperref}
\usepackage{subfigure}
\usepackage{epstopdf}
\usepackage{verbatim} 
\usepackage{array}
\usepackage{booktabs}
\usepackage{color}

\def\beq{\begin{equation}}
\def\eeq{\end{equation}}

\def\bea{\arraycolsep .1em \begin{eqnarray}}
\def\eea{\end{eqnarray}}
\def\Tr{{\rm Tr}}
\newcommand{\step}{\vspace{.5em}}

\def\eps{\epsilon}

\def\eq#1{(\ref{#1})}

\def\s0#1#2{\mbox{\small{$ \frac{#1}{#2} $}}}
\def\0#1#2{\frac{#1}{#2}}

\def\grgl{\:\hbox to -0.2pt{\lower2.5pt\hbox{$\sim$}\hss}{\raise3pt\hbox{$>$}}\:}
\def\klgl{\:\hbox to -0.2pt{\lower2.5pt\hbox{$\sim$}\hss}{\raise3pt\hbox{$<$}}\:}

\newcommand \be {\begin{equation}}
\newcommand \ee {\end{equation}}
\newcommand \bed {\begin{displaymath}}
\newcommand \eed {\end{displaymath}}

\newcommand{\bit}{\begin{itemize}}
\newcommand{\eit}{\end{itemize}}

\usepackage{colortbl}

\definecolor{Gray}{gray}{0.85}
\definecolor{LightGray}{gray}{0.93}
\definecolor{LightGreen}{rgb}{0.88, 1, 0.88}
\definecolor{LightCyan}{rgb}{0.88,1,1}
\definecolor{LightRed}{rgb}{1, 0.85, 0.85}
\definecolor{LightRed}{rgb}{1, 0.88, 0.88}
\definecolor{LightYellow}{rgb}{1, 1, 0.85}
\definecolor{LightBlue}{rgb}{0.87, 0.94, 1}
\definecolor{white}{gray}{1}

\usepackage{array,mathtools,amssymb,booktabs}
\newcolumntype{C}{>{$}c<{$}}
\AtBeginDocument{
\heavyrulewidth=.16em
\lightrulewidth=.1em
\cmidrulewidth=.03em
\belowrulesep=.4ex
\belowbottomsep=0pt
\aboverulesep=.4ex
\abovetopsep=0pt
\cmidrulesep=\doublerulesep
\cmidrulekern=.5em
\defaultaddspace=.5em
}

\makeatletter
    \def\CT@@do@color{%
      \global\let\CT@do@color\relax
            \@tempdima\wd\z@
            \advance\@tempdima\@tempdimb
            \advance\@tempdima\@tempdimc
    \advance\@tempdimb\tabcolsep
    \advance\@tempdimc\tabcolsep
    \advance\@tempdima2\tabcolsep
            \kern-\@tempdimb
            \leaders\vrule
                    \hskip\@tempdima\@plus  1fill
            \kern-\@tempdimc
            \hskip-\wd\z@ \@plus -1fill }
            
\begin{document}

\title{Asymptotic safety guaranteed in supersymmetry}

\author{Andrew D.~Bond}

\author{Daniel F.~Litim}

\affiliation{\mbox{Department of Physics and Astronomy, U Sussex, Brighton, BN1 9QH, U.K.}}

\begin{abstract}
We explain how asymptotic safety arises in four-dimensional supersymmetric gauge theories. 
We provide asymptotically safe 
supersymmetric
gauge theories 
together with
their superconformal fixed points, $R$-charges, phase diagrams, and UV-IR connecting trajectories.
Strict perturbative control is achieved in a Veneziano limit. 
Consistency with unitarity and the $a$-theorem is   established. 
We  find that supersymmetry enhances the predictivity of asymptotically safe theories. 
\end{abstract}

\maketitle

{\it Introduction.---}
The discovery of asymptotic freedom for non-abelian gauge theories in 1973 has initiated a new era in particle physics \cite{Gross:1973id,Politzer:1973fx}. Asymptotic freedom  explains why certain types of quantum field theories
such as the strong and weak sector of the Standard Model,
can be truly fundamental and predictive up to  highest energies. 
It implies that interactions are switched off asymptotically, and theories become free.
Asymptotic freedom constitutes a cornerstone in the Standard Model of particle physics, and continues to play an important role in the search for models beyond.

The  discovery of exact asymptotic safety for non-abelian gauge theories with matter \cite{Litim:2014uca,Bond:2016dvk,Bond:2017lnq}  has raised  substantial interest.  
Asymptotic safety explains how theories can be fundamental, predictive,  and {\it interacting} at highest energies  \cite{Wilson:1971bg}. Initially put forward as a scenario to quantize gravity \cite{Weinberg:1980gg,Reuter:1996cp,Litim:2003vp,
Falls:2013bv},  
asymptotic safety also arises in 
many other theories 
\cite{Bardeen:1983rv,
deCalan:1991km,Braun:2010tt,Wellegehausen:2014noa}. 
In particle physics,
 asymptotic safety offers 
intriguing new directions to ultraviolet (UV) complete the Standard Model
beyond the confines of asymptotic freedom 
\cite{Shaposhnikov:2009pv,Litim:2011cp,Bond:2017wut}.

  In this Letter, we  investigate whether asymptotic safety can be achieved in supersymmetric gauge theories.  In the language of the renormalisation group, asymptotic safety 
corresponds to an interacting 
UV fixed point for the running couplings \cite{Wilson:1971bg}.  Supersymmetry  modifies 
  fixed points and 
  the evolution of couplings 
because it links bosonic with fermionic degrees of freedom \cite{Martin:2000cr,Intriligator:2015xxa,Bond:2016dvk}. 
Additional constraints arise as bounds on the  superconformal $R$-charges  \cite{Intriligator:2003jj}  from both unitarity \cite{Mack:1975je} and  the $a$-theorem  \cite{Cardy:1988cwa,Osborn:1989td,Anselmi:1997am,Komargodski:2011vj}.
Hence, our task consists of finding supersymmetric gauge theories without asymptotic freedom, but  with viable interacting UV fixed points, and  in accord with all constraints. 
      
      One arena   in which   we may hope to find reliable answers is that of perturbation theory.  For sufficiently small couplings   \cite{Veneziano:1979ec},  the loop expansion and weakly interacting fixed points  are trustworthy  \cite{Bond:2016dvk}.
In this spirit, we  obtain  fixed points, phase diagrams, superconformal $R$-charges, and UV-IR connecting trajectories for supersymmetric gauge theories in a controlled setting.
Previously, this philosophy has been used 
successfully for proofs  of asymptotic safety in non-supersymmetric simple  \cite{Litim:2014uca} and semi-simple  \cite{Bond:2017lnq} gauge theories. \step

{\it The model.---}  
We consider a family of massless supersymmetric Yang-Mills theories in four space-time dimensions 
with product gauge group $SU(N_1)\otimes SU(N_2)$, coupled to
chiral superfields $(\psi,\chi,\Psi,Q)$ with flavour multiplicities $(N_F,N_F,1,N_Q)$. 
The main novelty is the use of a semi-simple gauge group as otherwise asymptotic safety cannot arise at weak coupling \cite{Bond:2016dvk,Martin:2000cr}. For each superfield we introduce a left- and right-handed copy with gauge charges 
as in Tab.~\ref{matter} to ensure the absence of gauge anomalies.
Also, viable models with asymptotic safety must have Yukawa couplings \cite{Bond:2016dvk}. Therefore, we allow for superpotentials of the form
\beq\label{W}
W= y\,\Tr\big[\psi_L\,\Psi_L\,\chi_L+\psi_R\,\Psi_R\,\chi_R\big]\,,
\eeq
where the trace sums over flavour and gauge indices. 
The superfields $Q$ are not furnished with Yukawa interactions.
The theory has a  global
$SU(N_F)_L\otimes SU(N_F)_R\otimes SU(N_Q)_L\otimes SU(N_Q)_R$   flavour and a $U(1)_R$ symmetry.
Moreover, the theory is renormalizable in perturbation theory and characterised by 
two gauge  couplings $g_1$ and $g_2$ and  the Yukawa coupling $y$, which we write as
\beq\label{alpha}
\begin{array}{rl}
\alpha_1=&
\displaystyle
\frac{N_1 \,g_1^2}{(4\pi)^2}\,, \quad\alpha_2=\frac{N_2 \,g_2^2}{(4\pi)^2}\,,
\quad
\alpha_y=
\displaystyle
\frac{N_1 \,y^2}{(4\pi)^2}\,.
\end{array}
\eeq
Sending  field multiplicities $(N_1, N_2, N_F, N_Q)$ to infinity while keeping their ratios fixed reduces the number of free parameters down to three, which we choose to be
\beq\label{Peps}
\begin{array}{rcl}
R&=&
\displaystyle
\0{N_2}{N_1}\,,
\displaystyle
\quad P=\0{N_1}{N_2}\0{N_Q+N_1+N_F-3N_2}{N_F+N_2-3N_1}\,,\\[2ex]
\eps&=&
\displaystyle
\0{N_F+N_2-3N_1}{N_1}\,.
\end{array}
\eeq
\noindent 
In the large-$N$ 
limit \cite{Veneziano:1979ec} the model parameters $(R,P,\eps)$ 
are continuous. We can always arrange to find \eq{Peps}  with
\beq\label{small}
1< R < 3\,, \quad P= {\rm finite}\,,\quad 0<|\eps|\ll 1\,.
\eeq
The smallness of $\eps$ ensures perturbative control in both gauge sectors \cite{Bond:2016dvk,Bond:2017lnq}, which is the  regime of interest for the rest of this work (the general case is discussed elsewhere \cite{BondLitim}).
This completes the definition of our models. 
\step

\begin{table}[b]
\begin{center}
\begin{tabular}{cc cc cc cc cc}
\toprule
\rowcolor{Yellow}
\ \ 
\bf Chiral superfields\ \ 
&$\bm{\psi_L}$ 
&$\bm{\psi_R}$ 
&$\bm{\Psi_L}$ 
&$\bm{\Psi_R}$ 
&$\bm{\chi_L}$ 
&$\bm{\chi_R}$ 
&$\bm{Q_L}$ 
&$\bm{Q_R}$ 
\\
\midrule
&& && && && \\[-3mm]
$\ \ \bm{SU(N_1)}\ \ $
&$\overline{\Box}$
&$
\Box$
&$\Box$
&$\overline{\Box}$
&1
&1
&1
&1
\\
&& && && && \\[-3mm]
\rowcolor{LightGray}
&& && && && \\[-3mm]
\rowcolor{LightGray}
$\bm{SU(N_2)}$
&1
&1
&$\Box$
&$\overline{\Box}$
&$\overline{\Box}$
&$\Box$
&$\overline{\Box}$
&$\Box$
\\[.4mm]
\bottomrule
\end{tabular}
\end{center}
 \vskip-.4cm
 \caption{\label{matter}Chiral superfields and their gauge charges.}
\end{table}

\begin{table*}
\begin{center}
\begin{tabular}{cGc Gc Gc G}
\toprule
\\[-4.mm]
\rowcolor{Yellow}
&&&&&&&\\[-3mm]
\rowcolor{Yellow}
\bf \ \ Fixed point\ \ 
&\bf \ \ \ G\ \ \
&
\ \ $\bm{{\rm BZ}{}_1}\ \ $ 
&
\ \ $\bm{{\rm BZ}{}_2}\ \ $ 
&
\ \ \ \ \ $\bm{{\rm GY}{}_1}\ \ \ \ \ $ 
&
\ \ \ $\bm{{\rm GY}{}_2}\ \ \ $ 
&
\ \ \ $\bm{{\rm BZ}{}_{12}}\ \ \ $ 
&
$\bm{{\rm GY}{}_{12}}$ 
\\
\midrule
$\bm{\alpha_1^*}$ 
&0
&$\ \ \ -\s0{\eps}{6}\ \ \ $
&0 
&$\ \ \ \s0{-\eps}{2 (3 - 3 R + R^2)}$\ \ \  
&0
&$\0{P R - 3}{16} {\eps}$
&$
\0{3  - 4  R - 2  P R^2 + P R^3}{(R-1) (9 - 8 R + 3 R^2)}\0{\eps}{2}$  \\[1mm]
$\bm{\alpha_2^*}$ 
&0
&0
&$\ \ \ -\0{P\eps}{6}\ \ \ $
&0
&\ \ \ $\0{-P R}{4 R-3}\0{\eps}{2}\ \ \ $ 
&\ \ \ $\0{1 - 3 P R}{16 R}\eps$\ \ \ 
&$
\ \ \0{R-2 - 3 P R + 3 P R^2 - P R^3}{ (R-1) (9 - 8 R + 3 R^2)}\0\eps{2}\ \ $ \\[1mm]
$\bm{\alpha_y^*}$ 
&0
&0
&0 
&$\s012\alpha_1^*$
&$\s012\alpha_2^*$
&0
&$\s012(\alpha_1^*+\alpha_2^*)$  \\[.4mm]
\bottomrule
\end{tabular}
\end{center}
 \vskip-.5cm
 \caption{\label{FPs}The Gaussian (G) and all
 Banks-Zaks (BZ) and gauge-Yukawa (GY) fixed points to leading order in $\eps$.
}
 \vskip-.3cm
\end{table*}

{\it Superconformal fixed points.---} 
The running of couplings 
 is controlled by  the beta functions $\beta_i=d\alpha_i/d\ln\mu$, with $\mu$ denoting the RG momentum scale. 
To find  accurate fixed points, 
we must minimally retain  terms up to two loop in the gauge and one loop in the Yukawa beta functions \cite{Bond:2016dvk}.
Using the results  of \cite{Einhorn:1982pp,Machacek:1983tz} and suppressing subleading terms in $\eps$, we find
\bea
	\beta_1 &=& 
	\displaystyle
	2\alpha_1^2\big[\ \eps + 6\alpha_1 + 2R \alpha_2 
	- 4R(3 - R)\alpha_y\big]
\nonumber
\,,\\
\label{beta12y}
	\beta_2 &=& 
		\displaystyle
2\alpha_2^2\Big[P\eps + 6\alpha_2 + \02R\alpha_1  
		- \04R(3 - R)\alpha_y\Big]
\,,\\
		\nonumber
	\beta_y&=& 
		\displaystyle
4\alpha_y\big[2\alpha_y 
- \alpha_1 - \alpha_2\big]\,.
\eea
Anomalous dimensions of the superfields are given by
\beq\label{anom}
\begin{array}{rcl}
 \gamma_\Psi&=&(3-R)\alpha_y - \alpha_1-\alpha_2\,,\\[.5ex]
\gamma_\psi&=&R\,\alpha_y - \alpha_1\,,\\ [.5ex]
\gamma_\chi&=&\alpha_y-\alpha_2\,,\\[.5ex]
\gamma_Q&=&-\alpha_2\,,
\end{array}
\eeq
up to  corrections of order ${\cal O}(\eps\,\alpha,\alpha^2)$.  
The simultaneous vanishing of \eq{beta12y} implies fixed points and scale invariance.
Besides the free Gaussian (G), the model has weakly coupled  fixed points $\alpha^*$ of order $\eps$.
These are either of the Banks-Zaks (BZ) or gauge-Yukawa (GY) type, depending on whether the Yukawa coupling is free 
or interacting 
\cite{Bond:2016dvk}. 
We find partially interacting Banks-Zaks  (${\rm BZ}_1, {\rm BZ}_2$)  and gauge-Yukawa (${\rm GY}_1, {\rm GY}_2$)  fixed points, and fully interacting ones  (${\rm BZ}_{12}, {\rm GY}_{12}$), all summarised in Tab.~\ref{FPs}. 
Results are exact to the leading order in $\eps$, with higher loop orders only correcting subleading terms.
We also note that \eq{beta12y}, \eq{anom},
and fixed points, are universal and RG scheme independent at weak coupling \cite{Litim:2014uca,Bond:2016dvk}. 

At superconformal fixed points, our models display a global and anomaly-free $U(1)_R$ symmetry.  In terms of 
the superfield anomalous dimensions 
\eq{anom}, the $R$-charges  
(not to be confused with the parameter $R$) 
read 
\beq\label{Ri}
R_i=2\left(1+\gamma^*_i\right)/3\,.
\eeq
Non-perturbative expressions for the $R$-charges are found using the method of $a$-maximisation \cite{Intriligator:2003jj}. For small couplings, findings agree with \eq{anom}, \eq{Ri} and deviate mildly from Gaussian values,  in accord with  unitarity \cite{Mack:1975je}.

Asymptotic freedom of \eq{beta12y} is guaranteed for
$P>0> \eps$.
Then, 
all three couplings \eq{alpha} are marginally relevant at the Gaussian UV fixed point. The set of asymptotically free trajectories 
is characterised by three free parameters, the initial values $0<\delta\alpha_i(\Lambda)\ll 1$ at the high scale $\Lambda$.
Some or all interacting fixed points of Tab.~\ref{FPs} arise 
within specific parameter ranges \eq{Peps} and take the role of   IR fixed points. Trajectories  run either towards a regime with strong coupling and confinement, or terminate at
a superconformal IR fixed point.  
By and large, this is very similar to the generic behaviour of 
asymptotically free  non-supersymmetric gauge theories  \cite{Bond:2017lnq}.

{\it Asymptotic safety.---} Next, we turn to  
 regimes \eq{Peps} where asymptotic freedom is lost, starting with
\beq \label{AS}
P<0<\eps\,.
\eeq
Clearly, the Gaussian has ceased to be the UV fixed point for the full theory and one might wonder whether its role is taken over by one of the interacting fixed points in Tab.~\ref{FPs}. Available candidates in the regime \eq{AS} are 
 BZ${}_2$, 
 GY${}_2$, and 
 GY${}_{12}$.
At the partially interacting BZ${}_2$, only the Yukawa term \eq{W}  is a relevant perturbation. The theory  becomes interacting in $\alpha_2$ and $\alpha_y$, yet $\alpha_1$ remains switched off at all scales. From the eigenvalue spectrum  
we  learn that GY${}_{12}$, once it exists, is IR attractive in all couplings.
 Hence, neither the Gaussian, nor BZ${}_2$, nor GY${}_{12}$ qualify as UV fixed points. 
\begin{figure}[b]
\begin{center}
\includegraphics[scale=.85]{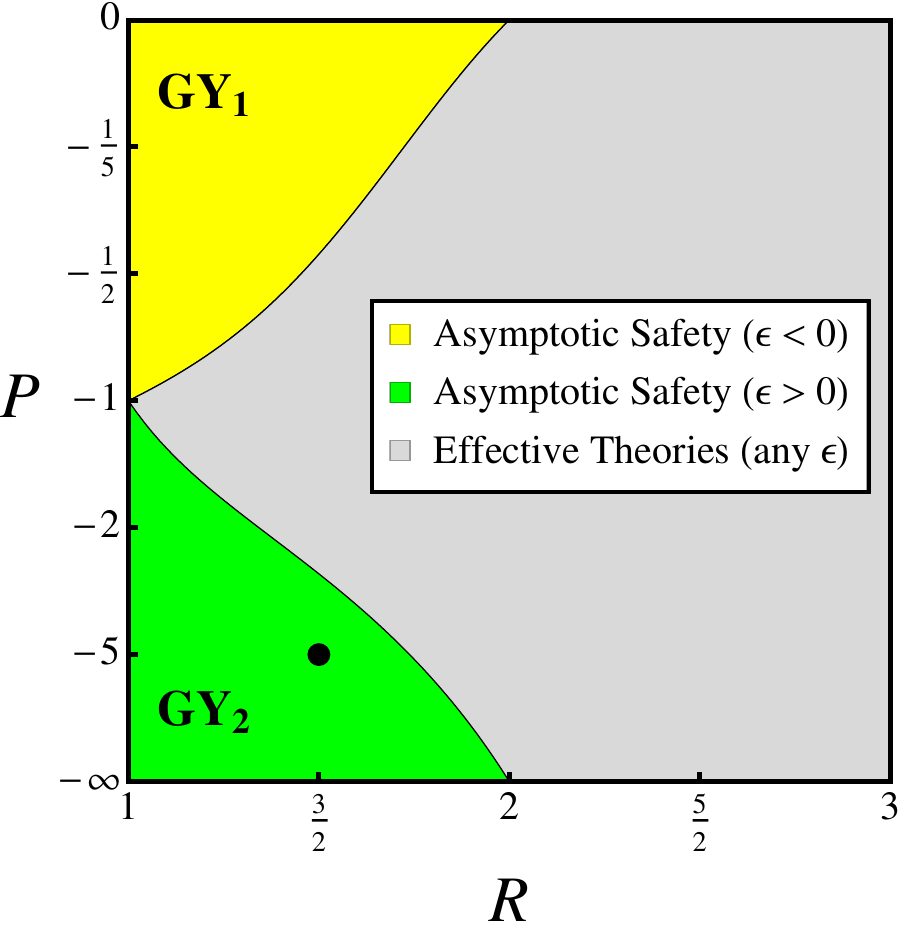}
\vskip-.2cm
\caption{Phase space for asymptotic safety, showing the parameter regions \eq{green} and \eq{yellow}. Models in the gray-shaded area are UV incomplete. $P$-axis is scaled as $P/(1-P)$ for better display. The full dot indicates the example in Figs.~\ref{pPDsusy},~\ref{pUVIR}.
}\label{pAllsusy}
\vskip-.2cm
\end{center}
\end{figure}
A new effect occurs at GY${}_2$. While $\alpha_2$ and $\alpha_y$ are irrelevant in its vicinity \cite{Bond:2016dvk}, the relevancy of $\alpha_1$ now  depends on the magnitude of $\alpha^*_2$ and $\alpha^*_y$ at GY${}_2$.  We find
\beq\label{beta1green}
\begin{array}{rcl}
\beta_1\big|_{{\rm GY}_2}
&=&
-B_{1,\rm eff}\,\alpha_1^2+{\cal O}(\alpha_1^3)\,,\\[1ex]
B_{1,\rm eff}&=&
\displaystyle
-2\eps+2\eps\,{P}/{Q_1}\,,
\end{array}
\eeq
with  $Q_1(R)=(4R-3)/(R^3-2R^2)$.
The first term in $B_{1,\rm eff}$ is the conventional one loop coefficient. It is negative in the regime \eq{AS} and documents the irrelevancy of $\alpha_1$ at the Gaussian. 
The second term  is sourced through the fixed point GY${}_2$. 
Most notably, the  sign of $B_{1,\rm eff}$ is positive provided that
\beq\label{green}
 P<Q_1<0\,,\quad 1<R<2\,,\quad \eps>0\,,
 \eeq
thereby turning $\alpha_1$ into a relevant coupling. We emphasize that the Yukawa term \eq{W} is crucial to achieve $B_{1,\rm eff}>0$; without it, the required change of sign would be  impossible  \cite{Bond:2016dvk}.
In other words, while $\alpha_1$ is IR free close to the Gaussian or BZ${}_2$ fixed points, it has become UV free close to the GY${}_2$ fixed point.  It is precisely for this reason that the gauge-Yukawa fixed point GY${}_2$ takes the role of an asymptotically safe UV fixed point with one marginally relevant and two irrelevant directions. 

The same mechanism is operative once
$P,\eps <0$,
where $\alpha_1$ 
and $\alpha_2$ 
have interchanged their roles.
Near GY${}_1$, the
effective one-loop coefficient for $\alpha_2$ reads $B_{2,\rm eff}=2 (Q_2-P)\, \eps$, with $Q_2=(R-2)/(R^3-3R^2+3R)$. Consequently, $\alpha_2$ becomes a relevant coupling for
\beq\label{yellow}
Q_2<P<0\,,\quad 1<R<2\,,\quad \eps<0\,,
\eeq
thereby promoting GY${}_1$ to an UV fixed point.  
As soon as both gauge sectors are destabilised $(P,\eps>0)$, no  fixed point  other than the IR attractive Gaussian can arise. Theories are UV incomplete and must be viewed as effective. Fig.\ref{pAllsusy} summarises our results once $P<0$, also indicating the parameter regions \eq{green} and \eq{yellow} with exact asymptotic safety.\step

\begin{figure}
\begin{center}
\includegraphics[scale=.22]{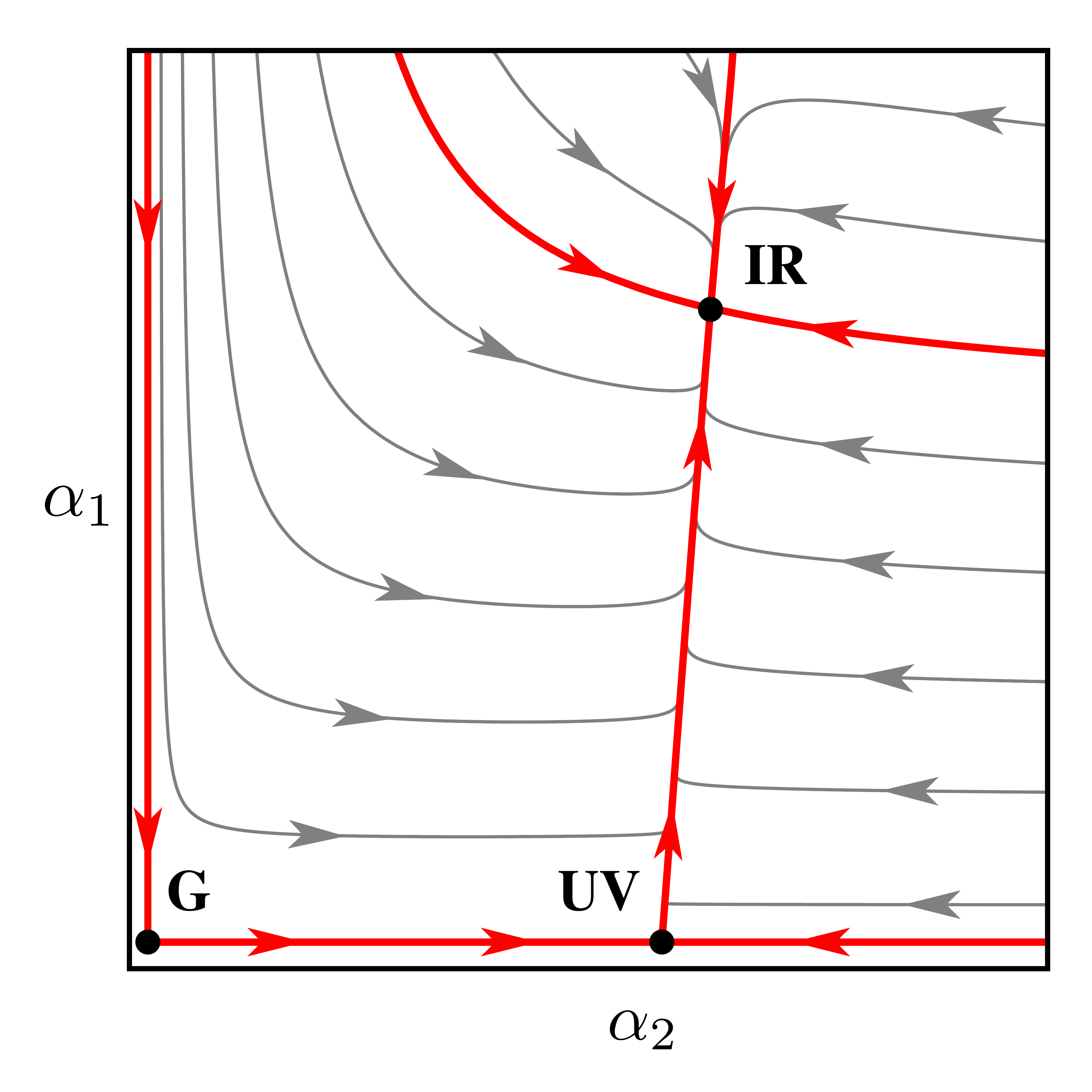}
\vskip-.2cm
\caption{
Phase diagram with asymptotic safety for supersymmetry
($P=-5, R=\032$, $\eps=\01{1000}$; Fig.~\ref{pAllsusy}) projected onto $\alpha_y=\s0{\alpha_1+\alpha_2}{2}$. Trajectories are pointing towards the IR.  Notice that $\alpha_1$ is destabilised and asymptotic freedom is absent.
Dots show the Gaussian, the UV and the IR fixed points.
Also shown are separatrices (red) and sample trajectories (gray).}\label{pPDsusy}
\end{center}
\vskip-.6cm 
\end{figure}
\begin{figure}
\begin{center}
\includegraphics[scale=.185]{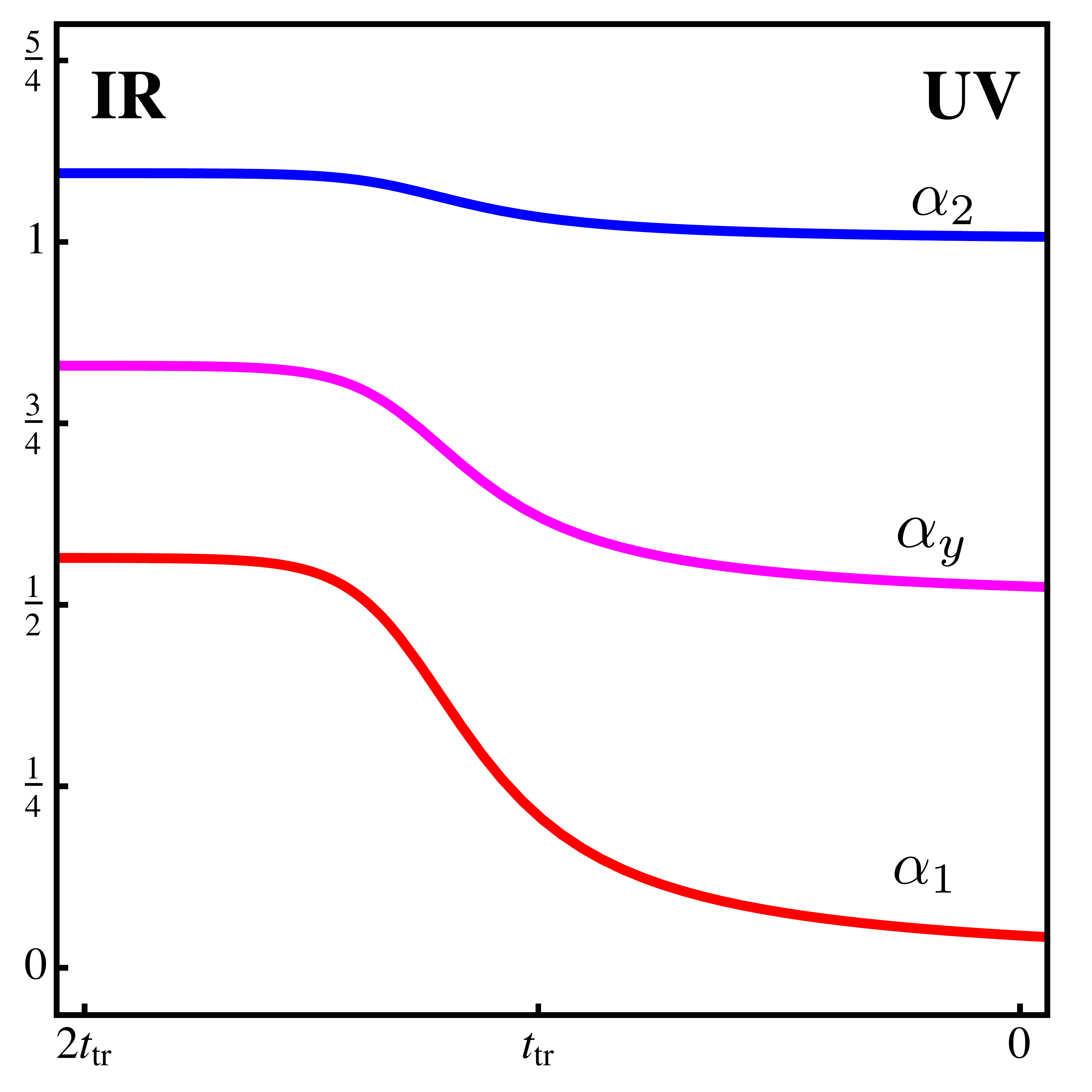}
\caption{\label{pUVIR}
The running couplings $\alpha_i(t)$ in units of RG time $t=\ln(\mu/\Lambda)$ along the 
separatrix from the UV to the IR fixed point.
Parameters as in Fig.~\ref{pPDsusy}. All couplings in units of $\alpha_{2,\rm UV}^*$ with $t_{\rm tr}=\ln (\mu_{\rm tr}/\Lambda)$ and $\Lambda$ the high scale, see \eq{mutr}.}
 \end{center}
 \vskip-.65cm
\end{figure}

{\it From the UV to the IR.---} 
At either of the superconformal UV fixed points, the elementary ``quarks'' and ``gluons'' are unconfined and appear as interacting (free) massless particles in one (the other) gauge sector.
The free gauge sector acts as a marginally relevant perturbation which
drives the theory away from the UV fixed point. The corresponding phase diagram in the regime \eq{green} 
is shown in Fig.\ref{pPDsusy}.
It confirms that GY${}_2$, unlike the Gaussian,  is the unique UV fixed point.
Close to the UV fixed point, the critical surface of asymptotically safe trajectories running out of it
is given by
\beq\label{sep}
\begin{array}{rcl}
\alpha_1(\mu)&=&
\displaystyle
\frac{\delta\alpha_1(\Lambda)}{1
+B_{1,\rm eff}\,
\delta\alpha_1(\Lambda)\,\ln(\mu/\Lambda)}\,,
\\[2ex]
\alpha_2(\mu)&=&
\displaystyle
\alpha^*_2+\0{2-R}{4R-3}\, 
\alpha_1(\mu)\,,
\\[2ex]
\alpha_y(\mu)&=&
\displaystyle
\alpha^*_y+\0{3R-1}{8R-6}\,
\alpha_1(\mu)\,.
\end{array}
\eeq
We emphasize that the theory has only one free parameter $\delta \alpha_1(\Lambda)\ll 1$ related to the relevant gauge coupling at the high scale $\Lambda$. 
  Both $\alpha_2$ and $\alpha_y$ have become irrelevant couplings and are strictly determined by $\alpha_1$.  (Similar expressions are found for the regime \eq{yellow}.) Dimensional transmutation leads to the RG invariant mass scale
\beq\label{mutr}
\mu_{\rm tr}
=\Lambda\exp\big[-{B_{1,\rm eff}\,\delta\alpha_1(\Lambda)}\big]^{-1}\,,
\eeq
which is independent of the high scale. 
It characterises the scale where couplings 
stop being controlled by the UV fixed point.
For RG scales  $\mu\ll \mu_{\rm tr}$, we observe  a cross-over into another superconformal fixed point  (GY${}_{12}$) governing the IR.
There, the elementary quarks and gluons of either gauge sector remain unconfined and appear as interacting massless particles, different from those observed in the UV.
Fig.~\ref{pUVIR} exemplifies the running of couplings 
from the UV to the IR.

The UV fixed point persists in the presence of mass terms for the chiral superfields. Once masses are switched on, with or without soft supersymmetry-breaking ones such as those for the ``gluinos'', they lead to decoupling \cite{Appelquist:1974tg} and low-energy modifications of the RG flow
 \eq{beta12y}. 
Then, UV safe trajectories may terminate in regimes with 
strong coupling and confinement in the IR, with or without softly broken supersymmetry.
\step

{\it Asymptotic safety and the a-theorem.---} 
We are now in a position to establish consistency 
with  a more formal aspect of the renormalisation group  known as the $a$-theorem 
\cite{Cardy:1988cwa,Osborn:1989td,Anselmi:1997am,Komargodski:2011vj}.
It states that  the central
charge $a=\frac{3}{32}\left[2d_G+\sum_i(1-R_i)(1-3(1-R_i)^2)\right]$
 \cite{Anselmi:1997am}, must be a decreasing function along 
RG trajectories in any $4d$ quantum field theory ($d_G$ denotes the dimension of the gauge groups and $i$ runs over all chiral superfields). 
Using \eq{anom}, \eq{Ri}, and Tab.~\ref{FPs},
we find
\beq
\Delta a\equiv a_{\rm UV}-a_{\rm IR} >0
\eeq
on any of the UV-IR connecting trajectories 
in the  parameter ranges \eq{green}, \eq{yellow} shown in  Fig.~\ref{pAllsusy}. 
Had the IR limit been the Gaussian, validity of the $a$-theorem 
implies
strong coupling and large $R$-charges in the UV, at least  for some of the fields \cite{Anselmi:1997am,Martin:2000cr}. In our models, this implication is circumvented
because the IR 
is not free.
In fact, there is not a single trajectory flowing from the UV fixed point to the Gaussian,
Fig.~\ref{pPDsusy},  which again is in accord with the $a$-theorem $(a_{\rm UV}-a_{\rm G} <0)$.

\step

{\it Discussion.---} 
In supersymmetry, and for superpotentials of the form \eq{W} including mass  terms, the scalar potential is always a sum of squares of absolute values  \cite{Martin:1997ns}. Hence, the stability of the quantum vacuum  is automatic.  Also, 
a  fixed point for the gauge and Yukawa couplings  implies a fixed point for the scalar potential. Without supersymmetry, physicality of scalar  fixed points and vacuum stability
do not come by default \cite{Bond:2016dvk} and must be checked case by case \cite{Litim:2015iea,Bond:2017lnq}.

Also, without supersymmetry, at least one Yukawa coupling is required to help generate an interacting UV fixed point \cite{Bond:2016dvk}. Invariably, this reduces the number of fundamentally free parameters in the UV by at least one, thereby enhancing the predictive power \cite{Litim:2014uca}. In supersymmetry, asymptotic safety at weak coupling cannot arise with only a single gauge factor \cite{Martin:2000cr,Bond:2016dvk}. 
Then, as we have seen in  \eq{sep}, at least one of the Yukawa couplings together with at least one of multiple gauge couplings must be non-trivial in the UV, thereby reducing the number of free parameters by two. 
We conclude that supersymmetry  additionally enhances the predictive power of asymptotic safety.
\step

We have shown that asymptotic safety is operative in supersymmetric gauge theories. Yukawa couplings continue to play a distinctive role at weak coupling, as they  do for asymptotic safety
 without supersymmetry \cite{Bond:2016dvk}. 
Explicit examples with superpotential \eq{W} and matter content as in Tab.~\ref{matter} are provided, including the phase space (Fig.~\ref{pAllsusy}) and phase diagram (Fig.~\ref{pPDsusy}).
Results are  consistent with unitarity and the $a$-theorem. 
Our construction makes it
clear that  asymptotic safety
exists in supersymmetry beyond the models discussed here.
It is interesting to include more gauge groups, expand Yukawa sectors, switch on mass terms, and explore the potential for asymptotically safe supersymmetric model building.\step\step

{\it Acknowledgements.---}
This work is supported by the Science and Technology Research Council (STFC) under the Consolidated Grant [ST/G000573/1] and by an STFC studentship.

\bibliographystyle{apsrev4-1}
\bibliography{bib_DFL}

 \end{document}